\documentclass[11pt,reqno]{amsart}
\usepackage[foot]{amsaddr}
\usepackage{graphicx}
\usepackage{amsthm,amssymb,amsmath,calc,eucal,ifthen}
\usepackage{amscd}
\usepackage[all,arc,curve,color,frame,matrix,arrow]{xy}
\usepackage{xcolor}
\usepackage{pgf,tikz,pgfplots}
\pgfplotsset{compat=1.15}
\usepackage{mathrsfs}
\usetikzlibrary{arrows}
\numberwithin{equation}{section}

\newtheorem{theorem}{Theorem}[section]
\newtheorem{lemma}[theorem]{Lemma}
\newtheorem{proposition}[theorem]{Proposition}

\theoremstyle{remark}
\newtheorem{remark}[theorem]{Remark}
\newtheorem{example}[theorem]{Example}

\newtheoremstyle{rmdefinition}{}{}{\upshape}{}{\bfseries}{.}{ }{}

\theoremstyle{rmdefinition}
\newtheorem{definition}[theorem]{Definition}

\newcommand{\be}[1]{\begin{equation}\label{#1}}
\newcommand{\ee}{\end{equation}}
\newcommand{\beqa}{\begin{eqnarray}}
\newcommand{\eeqa}{\end{eqnarray}}

\newcounter{tmpc}
\newlength{\tmplenght}
\newlength{\tmplenghta}
\newlength{\tmplenghtb}
\newlength{\tmplenghtc}

\begin{document}

\title[Set-indexed random fields]{Set-indexed random fields and algebraic Euclidean quantum field theory}

\author{Svetoslav Zahariev}
\address{MEC Department, LaGuardia Community College of The City University of New York, 31-10 Thomson Ave.,Long Island City, NY 11101, U.S.A.}
\email{szahariev@lagcc.cuny.edu}
\maketitle

\vspace{-7pt}

\vspace{-5pt}
\begin{abstract} We present a construction of non-Gaussian Borel measures on the space of continuous functions defined on the space of all balls in Euclidean space of arbitrary dimension. These measures induce nets of operator algebras satisfying the Haag-Kastler axioms of algebraic quantum field theory and may be interpreted as (nonlinear) continuous transformations of the free scalar massive Euclidean quantum field.
\end{abstract}

\section{Introduction}
In the axiomatic approach to quantum field theory (QFT) due to A. Wightman quantum fields are represented by point-like localized objects, namely, operator-valued distributions on Minkowski space. Consequently, several low-dimensional scalar QFT models have been constructed in the Euclidean framework as probability measures on the space of distributions on Euclidean spacetime. The passage to relativistic QFT is provided by the well-known Osterwalder-Schrader reconstruction theorem. 

The algebraic approach to QFT in which quantum fields are represented by nets, i.e. flabby pre-cosheaves, of operator algebras, transcends the necessity of using point-like localized objects. It is therefore natural to conjecture that models of algebraic QFT may be constructed from measures on more general topological vector spaces equipped with the structure of a flabby pre-sheaf over the Euclidean spacetime $\mathbb{R}^{d}$. It this article, we shall describe such a construction for any $d>1$ utilizing probabilistic methods and a version of the axioms of algebraic Euclidean QFT developed by D. Schlingemann in \cite{Sc}.

More specifically, we shall first associate to the free scalar massive Euclidean quantum field  a Gaussian measure  on the space of all functions from  the set of all balls in $\mathbb{R}^{d}$ to the real numbers. Second, employing well-known results on the sample path continuity of Gaussian random fields due to R. Dudley, we shall show that the latter measure can be promoted to a Borel measure on the space of all continuous functions defined on the space of all balls in $\mathbb{R}^{d}$. Finally, applying continuous (nonlinear) transformations to this measure, we shall obtain non-Gaussian measures on the same space and prove that these measures induce structures that satisfy the above mentioned axioms of algebraic Euclidean QFT.

While the non-Gaussian measures so obtained do not appear to bear direct relation to physically relevant QFT models, in a forthcoming paper we shall employ our method to construct interacting scalar QFT models such as the $\phi^4$-model via setting up and solving appropriate semilinear stochastic partial differential equations driven by an appropriately mollified white noise. Since we are dealing with probability measures defined on spaces of functions rather than distributions, such   stochastic equations would involve ordinary rather than generalized random fields.

This article is organized as follows. In Section \ref{prelimsect} we present some facts from the theory of Gaussian random fields that will be used in the sequel. In Section \ref{constrsetindsec} we discuss a general construction of Gaussian random fields indexed by the space of balls in $\mathbb{R}^{d}$ and their continuity properties. The axioms of algebraic Euclidean QFT as well as the relevant reconstruction theorem are briefly reviewed in Section \ref{axealgeusec}. Section \ref{fromranfisec} is dedicated to the passage from random fields indexed by balls in $\mathbb{R}^{d}$ to objects satisfying the axioms of algebraic Euclidean QFT over $\mathbb{R}^{d}$. Finally, in Section \ref{exatransmessec} we construct the non-Gaussian measures representing transformations of the mollified scalar massive free Euclidean quantum field.

\section{Preliminaries on Gaussian random fields}\label{prelimsect}
In this section we review some basic facts from the theory of Gaussian random fields.
In what follows we have fixed a probability space $(\Omega,\mathcal{F},\mu)$, where $\Omega$ is a set, $\mathcal{F}$ is a $\sigma$-algebra of subsets of $\mathcal{F}$, and $\mu$ is a probability measure on  $\mathcal{F}$.

Let $T$ be an arbitrary set. By a {\em random field} indexed by $T$ we shall mean a map $X$ from $T$ to the space of the real-valued random variables on $(\Omega,\mathcal{F},\mu)$. Given $t\in T$, we shall write $X_{t}$ for the random variable which is the image of $t$ under $X$. We say that $X$ is {\em  Gaussian} if all $X_{t}$ are centered Gaussian random variables, i.e., the image/pushforward of $\mu$ under $X_{t}$ is a centered Gaussian measure on $\mathbb{R}$ for all $t \in T$. In the remainder of this section we assume that all random fields considered consist of random variables in $L^2(\Omega,\mathcal{F},\mu)$, the space of square-integrable real-valued random variables on $(\Omega,\mathcal{F},\mu)$.

Recall that a function $C: T \times T \rightarrow \mathbb{R}$ is called {\em positive definite} if for all finite sequences $t_{1},\ldots,t_{n} \in T$ and $\lambda_{1},\ldots,\lambda_{n} \in \mathbb{R}$ one has
$$ \sum_{i,j=1}^{n}\lambda_{i}\lambda_{j}C(t_{i},t_{j})\geq 0.$$

The covariance of a random field $X$ defined via
$$\mathrm{Cov}(X)(t_{1},t_{2})=\langle  X_{t_{1}}, X_{t_{2}}\rangle_{L^2}, \quad t_{1}, t_{2} \in 
T,$$
is clearly positive definite and symmetric. 

Conversely, one can construct a Gaussian random field indexed by $T$ from any given symmetric positive definite function on $T  \times T$. Let us write $\mathbb{R}^{T}$ for the set of all functions from $T$ to $\mathbb{R}$ and $\mathcal{T}$ for the smallest $\sigma$-algebra on  $\mathbb{R}^{T}$ with respect to which the canonical projections from $\mathbb{R}^{T}$ to $\mathbb{R}$ are measurable.
\begin{theorem}\label{exigafie} Let $C: T \times T \rightarrow \mathbb{R}$ be positive definite and symmetric.
	
(1) Then there exists a centered Gaussian probability measure $\mu_{C}$ on $(\mathbb{R}^{T},\mathcal{T})$ and a Gaussian random field $X^{C}$ on $(\mathbb{R}^{T},\mathcal{T},\mu_{C})$ indexed by $T$ whose covariance is $C$.

(2) Let $\psi: T \rightarrow T$ be any map. Then the induced map $\bar{\psi}: \mathbb{R}^{T} \rightarrow \mathbb{R}^{T}$ is $\mathcal{T}$-measurable and one has
$$ \bar{\psi}_{\bullet}\mu_{C}=\mu_{C\circ (\psi\times \psi)},$$
where $\bar{\psi}_{\bullet}\mu_{C}$ is the image of $\mu_{C}$ under $\bar{\psi}$ and $\mu_{C\circ (\psi\times\psi)}$ is the Gaussian measure constructed in Part (1) from the covariance $C\circ (\psi\times \psi)$.
\end{theorem}
\begin{proof} This is well-known, see e.g. \cite[Theorem 8.2]{Ja}. We briefly describe the construction as it will be needed in the sequel. 
	
For every finite subset $S$ of $T$ one considers the centered Gaussian probability measure $\mu_{S}$ on $\mathbb{R}^{S}$ with covariance matrix $C_{ij}=C(t_{i},t_{j}), t_{i}\in S$. The Kolmogorov extension theorem (see e.g. \cite[Theorem 14.36]{Kl}) implies that the collection of these measures extends to a unique measure $\mu_{C}$ on $(\mathbb{R}^{T},\mathcal{T})$. The Gaussian random field $X^{C}$ is then simply the canonical field associated with the probability space $(\mathbb{R}^{T},\mathcal{T},\mu_{C})$, i.e., one sets
$X^{C}(t)(f)=f(t)$ for any $f \in \mathbb{R}^{T}$.
\end{proof}

We next assume that $T$ is a topological space and discuss certain continuity properties of random fields indexed by $T$ on a probability space $(\Omega,\mathcal{F},\mu)$. We shall call a random field $X$ {\em continuous} if it is continuous as a map from $T$ to $L^2(\Omega,\mathcal{F},\mu)$ considered with the norm topology. We say that $X$  {\em has continuous sample paths} if the real-valued function 
given by $ t \ni T \mapsto X_{t}(\omega)$ is continuous for almost all $\omega \in \Omega$ with respect to $\mu$.

Now let $d$ be a pseudo-metric on $T$ and assume that $T$ is compact with respect to the topology induced by $d$. We write $N_{d}(T,\varepsilon)$ for the minimum number of balls of radius $\varepsilon$ whose union covers $T$ and set
$$J(T,d)=\int_{0}^{\mathrm{diam}(T)}(\log N_{d}(T,\varepsilon))^{1/2}d\varepsilon,$$
where $\mathrm{diam}(T)$ stands for the diameter of $T$.
Given a Gaussian random field $X$  indexed by $T$, we define the {\em canonical pseudo-metric}  $d_{X}$ on $T$ associated to $X$ via 
$$ d^{2}_{X}(t_{1},t_{2})=\int_{\Omega} |X_{t_{1}}- X_{t_{2}}|^{2}d\mu .$$

Recall that a random field $\widetilde{X}$ is called a {\em modification} of $X$ if for every $t \in T$ one has $\widetilde{X}_{t}=X_{t}$ a.s. with respect to $\mu$. A proof of the following theorem due to Dudley may be found in \cite[Chapter 15, Section 4]{Ka}.

\begin{theorem}\label{dudtheo} If the integral $J(T, d_{X})$ is finite, the Gaussian random field $X$ has a modification with continuous sample paths on $(T, d_{X})$.
\end{theorem}
In what follows, we shall denote the Borel $\sigma$-algebra of a topological vector space $\mathcal{V}$ by $\mathcal{B}_{\mathcal{V}}$.

Now let $U$ be an open subset of $\mathbb{R}^{d}$ and let $X$ be a random field indexed by  $U$  that has a modification with continuous sample paths. We write $C(U)$ for the space of all continuous real-valued functions on $U$ equipped with the topology of uniform convergence on compact subsets of $U$. Proceeding as in \cite[Section 21.6]{Kl}, we shall explain how $\mu$ and $X$ induce a  probability measure on $(C(U), \mathcal{B}_{C(U)})$. 

By our assumption there exists $\bar{\Omega} \in \mathcal{F}$ with $\mu(\bar{\Omega})=1$ such that for every $\omega \in \bar{\Omega}$  the function given by $ x \ni U \mapsto X_{x}(\omega)$  is in $C(U)$. In other words, $X$ induces a map from $\bar{\Omega}$ to $C(U)$ which we denote by $\bar{X}$. Further, we write  $\bar{\mathcal{F}}$ for the restriction (trace) of the $\sigma$-algebra to $\bar{\Omega}$ and $\bar{\mu}$ for the restriction of $\mu$ to  $\bar{\mathcal{F}}$.

Let $\widetilde{X}_{x}: C(U) \rightarrow \mathbb{R}$ be the evaluation of a function at $x \in U$. It is standard (see e.g. \cite[p.125]{Ja}) that the $\sigma$-algebra on $C(U)$ generated by the evaluations $\widetilde{X}_{x}$ coincides with $\mathcal{B}_{C(U)}$. 
Since $ \widetilde{X}_{x}(\bar{X}(\omega))={X}_{x}(\omega)$ for all $x \in U$ and all $\omega \in \bar{\Omega}$, it follows that $\bar{X}$ is measurable map from $(\bar{\Omega},\bar{\mathcal{F}})$ to $(C(U), \mathcal{B}_{C(U)})$. Denoting by $\widetilde{\mu}$ the image of $\mu$ under $\bar{X}$, we obtain the canonical random field $\widetilde{X}$ on the probability space 
$(C(U), \mathcal{B}_{C(U)}, \widetilde{\mu})$  given by the evaluations $\widetilde{X}_{x}$.

\section{Constructing ball-indexed random fields}\label{constrsetindsec}
Let us denote the set of all open balls in $\mathbb{R}^{d}$ by $\mathrm{B}_{d}$. We identify
$\mathrm{B}_{d}$ with $\mathbb{R}^{d}\times (0,\infty)$ and equip it with  the Euclidean metric on $\mathbb{R}^{d}\times (0,\infty)$.

Given a positive selfadjoint bounded operator $A$ on $L^2(\mathbb{R}^{d})$, the space of the real-valued square integrable functions on $\mathbb{R}^{d}$, we construct a Gaussian measure and a Gaussian random field indexed by $\mathrm{B}_{d}$ associated to $A$ as follows. We write $b(x,r)$ for the open ball with radius $r$ centered at $x \in \mathbb{R}^{d}$, $1_{b(x,r)}$ for its indicator function, and set 

\begin{equation}\label{defofmolcova}
A(x,r,y,s):=v^{1/2}_{r}v^{1/2}_{s}\langle 1_{b(x,r)}, A1_{b(y,s)}\rangle_{L^2(\mathbb{R}^{d})},
\end{equation}
where $v_{r}$ is the reciprocal of the volume of a $d$-dimensional ball with radius $r$. The normalization constants in (\ref{defofmolcova}) are chosen so that $A(x,r,x,r)$ stays bounded as $r$ tends to 0 while at the same time it remains bounded away from 0 provided that the operator $A$ is bounded away from 0 (cf. Lemmas \ref{supboule} and \ref{exisoflimpo} below).

Since
$ \langle \hspace{1pt} \cdot \hspace{1pt} , A  \hspace{1pt} \cdot \hspace{1pt}  \rangle_{L^2(\mathbb{R}^{d})}$ is a symmetric positive definite bilinear form on $L^2(\mathbb{R}^{d})$,  $A(x,r,y,s)$ is symmetric and positive definite, considered as a function on $ \mathrm{B}_{d} \times \mathrm{B}_{d}$. Hence we can apply Theorem \ref{exigafie} to obtain a probability space $(\mathbb{R}^{\mathrm{B}_{d}}, \mathcal{B},\mu_{A})$ and a Gaussian random field $X^{A}$ on it.
\begin{proposition}\label{ahascontpath} The function $A(x,r,y,s)$ is continuous on $ \mathrm{B}_{d} \times \mathrm{B}_{d}$ and $X^{A}$ is a continuous random field. Moreover, if $A$ is bounded away from zero, $X^{A}$ has a modification which has continuous sample paths.
\end{proposition}
\begin{proof} 
	We observe that
\begin{multline}\label{firstesirxa}
\| X^{A}_{(x,r)}-X^{A}_{(y,s)}\|^{2}_{L^2(\mu_{A})}=A(x,r,x,r)+A(y,s,y,s)-2A(x,r,y,s)\\
=\|\sqrt{A} (v_{r}^{1/2}1_{b(x,r)}-  v_{s}^{1/2}1_{b(y,s)}) \|^{2}_{L^2(\mathbb{R}^{d})} \leq 
K\| (v^{1/2}_{r}1_{b(x,r)}-  v_{s}^{1/2}1_{b(y,s)}) \|^{2}_{L^2(\mathbb{R}^{d})}
\end{multline}
for some $K>0$. We fix a compact $V \subset \mathrm{B}_{d}$ and let $(x_{n},r_{n})$ be a sequence in $V$ converging to $(x_{0},r_{0})$. Since $v^{1/2}_{r_{n}}1_{b(x_{n},r_{n})}$ converges pointwise to $v^{1/2}_{r_{0}}1_{b(x_{0},r_{0})}$, we conclude by the Lebesgue dominated convergence theorem and (\ref{firstesirxa}) that $X^{A}$ is continuous on $V$ and hence on $\mathrm{B}_{d}$ as well. It follows (see e.g. \cite[Theorem 8.12]{Ja}) that  $A(x,r,y,s)$ is continuous on $ \mathrm{B}_{d} \times \mathrm{B}_{d}$.

Now consider the positive definite symmetric function on $ \mathrm{B}_{d} \times \mathrm{B}_{d}$ defined via
$$ W(x,r,y,s):=\langle 1_{b(x,r)}, 1_{b(y,s)}\rangle_{L^2(\mathbb{R}^{d})}
=\mathrm{vol}(b(x,r)\cap b(y,s)).$$
Applying Theorem \ref{exigafie}, we construct from  the function $W(x,r,y,s)$ a probability space $(\mathbb{R}^{\mathrm{B}_{d}}, \mathcal{B},\mu_{W})$ and a Gaussian random field $X^{W}$ on it, known as the ($d$-dimensional) Brownian sheet indexed by balls.

Let us compare the canonical pseudo-metrics $d_{X^{W}}$ and $d_{X^{A}}$  associated to $X^{W}$ and $X^{A}$ on an arbitrary compact subset $V$ of $\mathrm{B}_{d}$. We observe that $v_{r}$ is bounded from above, as well as bounded away from 0 when $(x,r)$ ranges within $V$, hence (\ref{firstesirxa}) implies
$$ d_{X^{A}}(b_{1},b_{2}) \leq K_{V,1} d_{X^{A}}(b_{1},b_{2}) $$
for all $b_{1},b_{2}\in V$ and some constant $K_{V,1}>0$. Similarly, since the operator $A$ is bounded away from zero, we find
$$ d_{X^{W}} (b_{1},b_{2})\leq K_{V,2} d_{X^{A}}(b_{1},b_{2}) $$
for some $K_{V,2}>0$, hence the pseudo-metrics $d_{X^{W}}$ and $d_{X^{A}}$ are strongly equivalent on $V$.

Next we note that by \cite[Theorem 1.4.7]{AT} (see also \cite[Theorem 8.13]{Du}) the integral $J(\mathrm{B}_{d}, d_{X^{^{W}}})$ is finite and so is, by the above equivalence, the integral $J(\mathrm{B}_{d}, d_{X^{^{A}}})$. Thus by Theorem \ref{dudtheo} the random field $X^{A}$ has a modification with continuous sample paths on $(V,d_{X^{A}})$. Since the compact $V$ is arbitrary and $\mathrm{B}_{d}$ is $\sigma$-compact, the same holds for $X^{A}$ on $(\mathrm{B}_{d},d_{X^{A}})$ (cf. \cite[p.13]{AT}). Finally, since convergence in $\mathrm{B}_{d}$ with respect to the Euclidean metric implies convergence with respect to the pseudo-metric $d_{X^{A}}$, we conclude that $X^{A}$ has a modification with continuous sample paths on $\mathrm{B}_{d}$ with respect to the Euclidean metric.
\end{proof}

Applying the construction at the end of Section  \ref{prelimsect} with $U=\mathrm{B}_{d}$ regarded as an open subset of $\mathbb{R}^{d+1}$, we obtain a probability space $(C(\mathrm{B}_{d}), \mathcal{B}_{C(\mathrm{B}_{d})}, \widetilde{\mu}_{A})$  and a canonical Gaussian random field $\widetilde{X}^{A}$ indexed by $\mathrm{B}_{d}$ on it.

\begin{example} When $A=I$, the identity operator, we shall refer to the associated random field $\widetilde{X}^{I}$ as the {\em mollified white noise}.
\end{example}

\begin{example}\label{exmolfref} When $A=(-\Delta+I)^{-1}$, where $\Delta$ is the Laplacian on $\mathbb{R}^{d}$, we shall refer to the associated random field $\widetilde{X}^{(-\Delta+I)^{-1}}$ as the {\em mollified free scalar Euclidean field of mass 1}.
\end{example}

\section{The axioms of algebraic Euclidean QFT}\label{axealgeusec}

In this section we present a version of the axioms and the reconstruction theorem of Algebraic Euclidean QFT  developed in \cite{Sc}.
We begin by introducing some notation. We write $(t, x_{1}, \ldots, x_{d-1})$ for the coordinates of $x \in \mathbb{R}^{d}$ and  set
$$\mathbb{R}^{d}_{+}=\left\{t, x_{1}, \ldots, x_{d-1} \in \mathbb{R}^{d}|t > 0\right\}, $$
$$\mathbb{R}^{d}_{-}=\left\{t, x_{1}, \ldots, x_{d-1} \in \mathbb{R}^{d}|t < 0\right\}, $$
$$\mathbb{R}^{d}_{0}=\left\{t, x_{1}, \ldots, x_{d-1} \in \mathbb{R}^{d}|t = 0\right\}. $$
Further, we write $\theta$ for the reflection
$$(t, x_{1}, \ldots, x_{d-1}) \mapsto (-t, x_{1}, \ldots, x_{d-1})$$
and denote by $\mathrm{E}_{d}$ the group of Euclidean motions in $\mathbb{R}^{d}$.


\begin{definition}\label{klrefpos} Let $\mathcal{H}$ be a Hilbert space and let $\{\mathfrak{S}(V)\}$ be a family $\{\mathfrak{S}(V)\}$ of closed subspaces of  $\mathcal{H}$ indexed by the open sets $V$ of $\mathbb{R}^{d}$ satisfying the following:

(1) (covariance) $U(g)\mathfrak{S}(V)U^{*}(g)=\mathfrak{S}(g \hspace{-1pt}\cdot \hspace{-1pt} V)$ for every $g \in \mathrm{E}_{d}$ and every open $V \subset \mathbb{R}^{d}$,

(2) (cyclicity) $\mathfrak{S}(\mathbb{R}^{d})=\mathcal{H}$,

(3) (continuity) Given an open $V \subset \mathbb{R}^{d}$, denote by $\Pi_{V}$ the orthogonal projection onto $\mathfrak{S}(V)$. Given an infinite sequence $V_{1}\subset V_{2} \subset V_{3} \subset \cdots$ of open subsets of $\mathbb{R}^{d}$ with $\cup_{i=1}^{\infty}V_{i}=V$, the sequence $\Pi_{V_{i}}$  convergences strongly to $\Pi_{V}$.

A strongly continuous unitary representation $\mathrm{E}_{d} \ni g \mapsto U(g)$ of $\mathrm{E}_{d}$ on $\mathcal{H}$ is called {\em reflection positive} with respect to $\{\mathfrak{S}(V)\}$ if one has $\langle h, U(\theta)h\rangle_{\mathcal{H}}\geq 0$
for every $h \in \mathfrak{S}(\mathbb{R}^{d}_{+})$. 
\end{definition}

The latter definition is motivated by the main result of \cite{KL} which shall be recalled below, in the proof of Theorem \ref{eclnetth}.

In what follows, by a {\em net of unital algebras} over a topological space $X$ indexed by a collection $\mathcal{V}$ of open subsets of $X$ we mean an assignment $V \mapsto \mathfrak{B}(V)$ of a unital algebra to every $V \in \mathcal{V}$ such that there is an inclusion of unital algebras $ \mathfrak{B}(V_{1}) \subset \mathfrak{B}(V_{2})$ whenever $V_{1} \subset V_{2}$. Unless otherwise specified, we shall assume that the  nets we consider are indexed by the collection of all open sets on $X$. (Such nets are, in modern terminology, nothing but flabby pre-cosheaves on $X$.)

We shall write $\mathrm{L(\mathcal{H})}$ for the algebra of all bounded operators on the Hilbert space $\mathcal{H}$.

\begin{definition}\label{aegftdefi} 
	A triple $(\mathcal{H},\mathfrak{A}, h_{0})$, where $\mathcal{H}$ is a complex Hilbert space, $\mathfrak{A}$ is a net of commutative von Neumann algebras of operators on $\mathcal{H}$  over  $\mathbb{R}^{d}$, and $h_{0}$ is a non-zero vector in $\mathcal{H}$, is called an {\em algebraic Euclidean QFT over $\mathbb{R}^{d}$} if the following holds.
	
	(1) The Hilbert space $\mathcal{H}$ is equipped with a strongly continuous unitary representation of $\mathrm{E}_{d}$ for which we write $\mathrm{E}_{d} \ni g \mapsto U(g)$.
	
	(2) The vector $h_{0}$ is of norm 1 and is $\mathrm{E}_{d}$-invariant.
	
	(3) The net $\mathfrak{A}$ satisfies:
	
	(a) (covariance) $U(g)\mathfrak{A}(V)U^{*}(g)=\mathfrak{A}(g \hspace{-1pt}\cdot \hspace{-1pt} V)$ for every $g \in \mathrm{E}_{d}$ and every open $V \subset \mathbb{R}^{d}$, where $\mathfrak{A}(V)$ stands for the algebra in the net $\mathfrak{A}$ corresponding to $V$;
	
	(b) (reflection positivity) The family of subspaces $\{\mathfrak{A}(V)h_{0}\}$ satisfies the assumptions of Definition \ref{klrefpos} and the representation $\mathrm{E}_{d} \ni g \mapsto U(g)$ is reflection positive with respect to this family.
	
	(c) (time-zero operators) We set 
	
	$$  \mathfrak{A}_{0}:=\{A \in \mathfrak{A}(\mathbb{R}_{+}^{d}) \cap \mathfrak{A}(\mathbb{R}_{-}^{d}): AU(\theta)=U(\theta)A \} $$
	and assume that  $\mathfrak{A}_{0}\neq \mathbb{C}$.
	
	\end{definition}

Next we describe the passage from algebraic Euclidean QFT to algebraic relativistic QFT satisfying the Haag-Kastler axioms (cf. \cite{HK} and \cite{A}). We denote by $\mathbb{M}^{d}$ the $d$-dimensional Minkowski space and by $\mathcal{P}^{\,d}$ the $d$-dimensional orthochronous Poincar\'e group. The following theorem is a reformulation of the main result of \cite{Sc}.

\begin{theorem}\label{eclnetth} Every  Euclidean algebraic QFT $(\mathcal{H},\mathfrak{A},h_{0})$ over $\mathbb{R}^{d}$ gives rise to a triple
	$(\widetilde{\mathcal{H}},\widetilde{\mathfrak{A}},\widetilde{h}_{0})$, 
	where $\widetilde{\mathcal{H}}$ is a complex Hilbert space, $\widetilde{\mathfrak{A}}$
	is a net of unital $C^*$-algebras acting on $\widetilde{\mathcal{H}}$  indexed by the double cones in  $\mathbb{M}^{d}$, $\widetilde{h}_{0}$ is a non-zero vector in $\widetilde{\mathcal{H}}$, and the following holds.
	
	(1) The Hilbert space $\widetilde{\mathcal{H}}$ is equipped with a strongly continuous unitary representation of $\mathcal{P}^{\,d}$ denoted by $\mathcal{P}^{\,d} \ni g \mapsto \widetilde{U}(g)$, which satisfies the {\em spectrum condition}:  the selfadjoint generators of the representation of the translation subgroup of $\mathcal{P}^{\,d}$ have spectrum contained in the closed forward light cone in $\mathbb{M}^{d}$.
	
	(2) (vacuum) The vector $\widetilde{h}_{0}$ is of norm 1 and is $\mathcal{P}^{\,d}$-invariant.
	
	(3) The net $\widetilde{\mathfrak{A}}$ satisfies:
	
	(a) (covariance) $\widetilde{U}(g) \widetilde{\mathfrak{A}}(\mathcal{O}) \widetilde{U}^{*}(g)= \widetilde{\mathfrak{A}}(g \hspace{-2pt}\cdot \hspace{-2pt} \mathcal{O}) $ for every $g \in \mathcal{P}^{\,d}$ and every double cone $\mathcal{O} \subset \mathbb{M}^{d}$.
	
	(b) (causality)  $[\widetilde{\mathfrak{A}}(\mathcal{O}_{1}), \widetilde{\mathfrak{A}}(\mathcal{O}_{2})]=\{0\}$ whenever $\mathcal{O}_{1}$ and $\mathcal{O}_{2}$ are space-like separated double cones in $\mathbb{M}^{d}$.
	
\end{theorem}

\begin{proof}
	
	We first construct a ``physical'' Hilbert space from $\mathcal{H}$ as follows (cf. \cite[ Section I.1]{KL}). Let $R_{\theta}= \Pi_{\mathbb{R}^{d}_{+}}U(\theta) \Pi_{\mathbb{R}^{d}_{+}}$ and set $\mathcal{H}_{0}=R_{\theta}(\mathcal{H})=
	R_{\theta}(\mathfrak{A}(\mathbb{R}^{d}_{+})h_{0})$. We define a sesquilinear form on $\mathcal{H}_{0}$ via
	$$ \langle R_{\theta}h_{1}, R_{\theta}h_{2} \rangle_{\theta}= \langle h_{1}, U(\theta) h_{2} \rangle_{\mathcal{H}}, \quad h_{1},h_{2} \in \mathfrak{A}(\mathbb{R}^{d}_{+})h_{0}.$$
	By the reflection positivity assumption (Definition \ref{aegftdefi} (3)(b)) the latter form is in fact an inner product; we denote the completion of $\mathcal{H}_{0}$ with respect to it by $\widetilde{\mathcal{H}}$. 
	
	Now applying the main result of \cite[ Section I]{KL} to the family of $\mathrm{E}_{d}$-covariant subspaces $\{\mathfrak{A}(V)h_{0}\}$, we obtain a strongly continuous unitary representation of $\mathcal{P}^{\,d}$ on $\widetilde{\mathcal{H}}$ satisfying the spectrum condition for which we write $\mathcal{P}^{\,d} \ni g \mapsto \widetilde{U}(g)$. Moreover, since $h_0$ is $\mathrm{E}_{d}$-invariant, $\widetilde{h}_{0}:=R_{\theta}h_{0}$ is non-zero and $\mathcal{P}^{\,d}$-invariant by construction.
	
	The $\mathcal{P}^{\,d}$-covariant net of $C^*$-algebras $\widetilde{\mathfrak{A}}$ acting on $\widetilde{\mathcal{H}}$ is then constructed out of $\mathfrak{A}_{0}$ exactly as in \cite[Section 3.2]{Sc} and its causality property follows from \cite[Theorem 3.7]{Sc}.
	\end{proof}


\section{From random fields to algebraic Euclidean QFT}\label{fromranfisec}

We fix a real locally convex space $\mathcal{V}$ such that $\mathcal{V}$ is a vector subspace of $C(\mathrm{B}_{d})$ and the topology on  $\mathcal{V}$ is not weaker than the topology of pointwise convergence on $C(\mathrm{B}_{d})$. We further fix a Borel probability measure $\mu$ on $\mathcal{V}$ and we write $\mathcal{H}_{\mu}$  for $L^{2}_{\mathbb{C}}(\mathcal{V}, \mathcal{B}_{\mathcal{V}}, \mu)$, the complexification of $L^2(\mathcal{V}, \mathcal{B}_{\mathcal{V}}, \mu)$.
 
Given an open $V \subseteq \mathbb{R}^{d}$, we set
$$\mathrm{B}_{d}(V)=\{(x,r)\in \mathrm{B}_{d}: b(x,r) \subseteq V\}. $$
Let $X$ be a random field indexed by $\mathrm{B}_{d}$ consisting of real-valued random variables in $\mathcal{H}_{\mu}$. We define $\mathfrak{A}_{X}(V)$ to be the von Neumann algebra generated by the operators
$$ \{ M_{X_{b(x,r)}} : (x,r) \in \mathrm{B}_{d}(V)\},$$
where $M_{X_{b(x,r)}}$  stands for the operator on $\mathcal{H}_{\mu}$ given by multiplication by the bounded measurable function $e^{iX_{b(x,r)}}$. Thus we obtain a net of von Neumann algebras over $\mathbb{R}^{d}$ which we denote by $\mathfrak{A}_{X}$. Several labels corresponding to elements of $\mathfrak{A}(\mathbb{R}^{2}_{+})$ are shown on Fig. 1 below.

\begin{figure}[!h]
\begin{tikzpicture}[line cap=round,line join=round,>=triangle 45,x=1cm,y=1cm,scale=0.5]
	\draw[thick,->] (-4,0) -- (6,0);
	\draw[thick,->] (0,-4) -- (0,6);
	\draw [line width=1pt] (2.6,3.34) circle (2.0153411621856976cm);
	\draw [line width=1pt] (2.58,0.22) circle (1.6201234520862908cm);
	\draw [line width=1pt] (0.8,-1.8) circle (0.7cm);
		
	\filldraw (2.6,3.34)  circle (2.5pt);
	\filldraw (2.58,0.22)  circle (2.5pt);
	\filldraw (0.8,-1.8)  circle (2.5pt);
		
	\draw [line width=2pt] (5.7,-0.7) node { $t$};
	\draw [line width=2pt] (0.8,5.7) node { $x_{1}$};	
	\draw [line width=2pt] (6,3) node {$\mathfrak{A}(\mathbb{R}^{2}_{+})$};
\end{tikzpicture}
\label{fig1}
\caption{}
\end{figure}	
	
Given a random field $X$ indexed by a set $T$ and a subset $V$ of $T$, we write $\mathrm{S}(X,V)$ for the $\mathbb{R}$-linear span of the random variables $\{X_{t}\}_{t \in V}$.
\begin{definition}\label{rposforanf} We say that a random field $X$ indexed by $\mathrm{B}_{d}$ is {\em reflection positive} if for each finite sequence $F_{1},\ldots, F_{n}$ in $\mathrm{S}(X,\mathrm{B}_{d}(\mathbb{R}^{d}_{+}))$ the matrix
	$$ S_{ij}=\int_{\mathcal{V}} e^{iF_{i}-i \theta \cdot F_{j}}d\mu $$
	is positive definite.
\end{definition}

We now equip $\mathrm{B}_{d}$ with the natural continuous action of $\mathrm{E}_{d}$ given by
$$ g \cdot b(x,r)=b(g\cdot x,r), \quad g \in \mathrm{E}_{d},$$
which induces actions of $\mathrm{E}_{d}$ on $\mathcal{V}$ and $\mathcal{H}_{\mu}$.  Below we write $1_{\mathcal{H}_{\mu}}$ for the constant function 1 in $\mathcal{H}_{\mu}$. 

\begin{theorem}\label{fromranfitoth} Assume that the measure $\mu$ is $\mathrm{E}_{d}$-invariant, the random field $X$ is reflection positive, continuous and $\mathrm{E}_{d}$-equivariant as a map from $\mathrm{B}_{d}$ to $\mathcal{H}_{\mu}$  and $\mathfrak{A}_{X}(\mathbb{R}^{d}) 1_{\mathcal{H}_{\mu}}=\mathcal{H}_{\mu}$.
	
Suppose further that for every $x \in \mathbb{R}^{d}$ there exists a sequence of positive numbers $r_{n}$ decreasing to 0 such that $e^{iX_{b(x,r_{n})}}$ converges in $\mathcal{H}_{\mu}$ to a non-constant random variable denoted by $e^{iX_{b(x,0)}}$ as $n \rightarrow \infty$.
	
Then the triple $ (\mathcal{H}_{\mu}, \mathfrak{A}_{X}, 1_{\mathcal{H}_{\mu}})$  is an algebraic Euclidean QFT over $\mathbb{R}^{d}$ in the sense of Definition \ref{aegftdefi}.
\end{theorem}
\begin{proof}
Since $\mu$ is $\mathrm{E}_{d}$-invariant and the action of $\mathrm{E}_{d}$ on $\mathcal{V}$ is continuous, the induced representation on $\mathcal{H}_{\mu}$ is unitary and strongly continuous, hence property (1) in Definition \ref{aegftdefi} holds. Property (3)(a) follows from the equivariance of $X$ and the fact that for any $g \in \mathrm{E}_{d}$ and any open $V \subseteq  \mathbb{R}^{d}$, $ b(x,r) \subset V$ implies that $g\cdot b(x,r) \subset g(V)$. 

Further, observe that property (3) in Definition \ref{klrefpos} follows directly from the construction of the net $\mathfrak{B}_{X}$, hence the reflection positivity of $X$ implies that property (3)(b) in Definition \ref{aegftdefi} is satisfied.

To show that property (3)(c) in Definition \ref{aegftdefi} holds, let $\mathfrak{A}_{X,0}$ stand for the algebra generated by the operators on $\mathcal{H}_{\mu}$ given by multiplication with $e^{iX_{b(x,0)}}$, where $x$ ranges within  $\mathbb{R}_{0}^{d}$. Now given $x \in \mathbb{R}_{0}^{d}$ and a sequence $r_{n}$ as in the statement of the theorem such that $e^{iX_{b(x,r_{n})}}\rightarrow e^{iX_{b(x,0)}}$, we can find a sequence $x_{n} \in \mathbb{R}_{+}^{d}$ such that $x_{n} \rightarrow x$ in $\mathbb{R}^{d}$ and $b(x_{n},r_{n}) \in  \mathrm{B}_{d}(\mathbb{R}_{+}^{d})$. The continuity of $X$ implies that $e^{iX_{b(x,r_{n})}}-e^{iX_{b(x_{n},r_{n})}}\rightarrow 0$ in $\mathcal{H}_{\mu}$, hence $\mathfrak{A}_{X,0} \subset 
\mathfrak{A}_{X}({\mathbb{R}_{+}^{d}})$. Similarly $\mathfrak{A}_{X,0} \subset 
\mathfrak{A}_{X}({\mathbb{R}_{-}^{d}})$ from which the desired property follows.
\end{proof}

\section{Example: Transformations of the mollified Euclidean free field}\label{exatransmessec}
It will be convenient for our purposes in this section to recast the notion of reflection positivity in terms of positive-definite kernels and probability measures. As a preparation, we discuss a finite-dimensional version of reflection positivity.

Let $\mu_{n}$ be a centered Gaussian probability measure on $\mathbb{R}^{2n}$ with covariance matrix $C^{\mu}_{ij}$. Let $\theta_{n}$ be an involutive permutation of the set $\{1,2,\ldots, 2n\}$ which maps $\{1,2,\ldots ,n\}$ to $\{n+1,n+2,\ldots ,2n\}$ and vice versa.
\begin{lemma}\label{findimreflpos}
If the matrix $\{C^{\mu}_{i,\theta_{n}(j)}\}_{i,j=1}^{n}$ is positive definite, then so is the  matrix
$$ S^{\mu}_{ij}= \int_{\mathbb{R}^{2n}}e^{ix_{i}-ix_{\theta_{n}(j)}}d\mu_{n}(x_{1},\ldots, x_{2n}),$$
where $i,j=1,\ldots, n$ and  $(x_{1},\ldots, x_{2n}) \in \mathbb{R}^{2n}$.
\end{lemma}
\begin{proof}
We observe that  $S^{\mu}_{ij}=S_{\mu_{n}}(e_{i}-e_{\theta_{n}(j)}) $,  where $S_{\mu_{n}}$ is the characteristic function of $\mu_{n}$ and $e_{1},\ldots, e_{2n}$ is the standard orthonormal basis of $\mathbb{R}^{2n}$. Thus the statement of the lemma is the finite-dimensional analogue of \cite[Theorem 6.2.2]{GJ}.
\end{proof}

Let $\mathcal{V}$ be as in Section \ref{fromranfisec} and assume further that $\mathcal{V}$ has the structure of the total space of a flabby pre-sheaf of locally convex spaces over $\mathrm{B}_{d}$, i.e., for every open subset $\mathrm{V}$ of $\mathrm{B}_{d}$ we are given a subspace $\mathcal{V}(\mathrm{V})$ of $\mathcal{V}$ and a continuous linear surjection $R_{\mathrm{V}}:\mathcal{V} \rightarrow 
\mathcal{V}(\mathrm{V})$ such that $R_{\mathrm{V}_{1}}=R_{\mathrm{V}_{1}}R_{\mathrm{V}_{2}}$
 whenever $\mathrm{V}_{1} \subset \mathrm{V}_{2}$ and one has $\mathcal{V}=\mathcal{V}(\mathrm{B}_{d})$.

Let $\mathrm{V}$ be an open subset of $\mathrm{B}_{d}$ and write $\mathcal{M}(\mathrm{V})$ for the vector space of all complex-valued bounded measurable functions  on $\mathcal{V}$ which have the following property: for each  $\Psi$ in $\mathcal{M}(\mathrm{V})$  there exists a bounded measurable function $\Psi'$ on 
$\mathcal{V}(\mathrm{V})$ such that $\Psi= \Psi' \circ R_{\mathrm{V}}$. We introduce the short-hand notation $\mathrm{B}_{d}^{+}=\mathrm{B}_{d}(\mathbb{R}^{d}_{+})$.

\begin{definition}\label{mesdefrpos}
A probability measure $\mu$ on $(\mathcal{V}, \mathcal{B}_{\mathcal{V}})$ is 
called {\em reflection positive} if
\begin{equation}\label{mesreflosot}
\int_{\mathcal{V}}\Psi (\overline{\theta \hspace{-2pt}\cdot \hspace{-2pt}\Psi})d\mu \geq 0
\end{equation}
for all $\Psi \in \mathcal{M}(\mathrm{B}_{d}^{+})$.
\end{definition}

There is a simple relation between the latter definition and Definition \ref{rposforanf}: If $\{X_{x}\}_{x \in \mathrm{B}_{d}}$ is a random field on $(\mathcal{V}, \mathcal{B}_{\mathcal{V}}, \mu)$ such that $e^{iX_{x}}\in \mathcal{M}(\mathrm{B}_{d}^{+})$ for all $x \in \mathrm{B}_{d}$ and $\mu$ is reflection positive, then it is easy to check that $\{X_{x}\}$ is reflection positive. 

Now let $A$ be a positive selfadjoint bounded operator on $L^2(\mathbb{R}^{d})$ and consider the Gaussian measures $\mu_{A}$ and $\widetilde{\mu}_{A}$ and the corresponding canonical random fields $X^{A}$ and $\widetilde{X}^{A}$ associated to $A$  that were constructed in Section \ref{constrsetindsec}.
\begin{lemma}\label{lemtwoparo} (1) Assume that $A$ is $\mathrm{E}_{d}$-invariant, i.e., $A$ commutes with the natural unitary representation of $\mathrm{E}_{d}$ on $L^2(\mathbb{R}^{d})$. Then the measure $\widetilde{\mu}_{A}$ is $\mathrm{E}_{d}$-invariant.
		
(2) Assume that $A$ is reflection positive, i.e., one has
$$\langle Af,  \theta f\rangle_{L^2}\geq 0$$
for all $f \in C^{\infty}_{c}(\mathbb{R}^{d}_{+})$.  Then the measure $\widetilde{\mu}_{A}$ is 
reflection positive.
\end{lemma}
\begin{proof}
(1)	The $\mathrm{E}_{d}$-invariance of $A$ implies
$$ A(g \cdot x,r,g \cdot y,s)=A(x,r,y,s)$$
for all $g \in \mathrm{E}_{d}, x,y \in \mathbb{R}^{d}, r,s \in (0,\infty)$. Thus by Theorem \ref{exigafie}(2) the measure $\mu_{A}$ is $\mathrm{E}_{d}$-invariant and therefore so is $\widetilde{\mu}_{A}$.  
	
(2) Given $(x_{1},r_{1}),\ldots, (x_{n},r_{n}) \in \mathrm{B}_{d}^{+}$, the reflection positivity of $A$ implies that the matrix
$$
A_{ij}^{\theta}:=A(x_{i},r_{i},\theta \cdot x_{j} ,r_{j})
$$ is positive definite.
\begin{figure}[!h]
	\begin{tikzpicture}[line cap=round,line join=round,>=triangle 45,x=1cm,y=1cm,scale=0.5]
		\draw[thick,->] (-6,0) -- (6,0);
		\draw[thick,->] (0,-5) -- (0,6.2);
		\draw [line width=1pt] (-4,4.52) circle (1.6cm);
		\draw [line width=1pt] (-1.7,1.8) circle (1.3cm);
		\draw [line width=1pt] (-3,-2.4) circle (2cm);
		\draw [line width=1pt] (4,4.52) circle (1.6cm);
		\draw [line width=1pt] (1.7,1.8) circle (1.3cm);
		\draw [line width=1pt] (3,-2.4) circle (2cm);
		
		\filldraw (-4,4.52)  circle (2.5pt);
		\filldraw (-1.7,1.8)  circle (2.5pt);
		\filldraw (-3,-2.4) circle (2.5pt);
		\filldraw (4,4.52)  circle (2.5pt);
		\filldraw (1.7,1.8)  circle (2.5pt);
		\filldraw (3,-2.4) circle (2.5pt);
		
		\node at   (4,4.1) {\tiny{$(x_{1},r_{1})$}};
		\node at   (1.7,1.4) {\tiny{$(x_{2},r_{2})$}};
		\node at   (3,-2.8) {\tiny{$(x_{n},r_{n})$}};
		
		\node at   (-4,4.1) {\tiny{$(\theta \hspace{-2pt}\cdot \hspace{-2pt} x_{1},r_{1})$}};
		\node at   (-1.7,1.4) {\tiny{$(\hspace{-1pt}\theta \hspace{-2pt}\cdot \hspace{-2pt} x_{2},\hspace{-1pt} r_{2})$}};
		\node at   (-3,-2.8) {\tiny{$(\theta \hspace{-2pt}\cdot \hspace{-2pt} x_{n},r_{n})$}};
		
		\draw [line width=2pt] (5.7,-0.7) node { $t$};
		\draw [line width=2pt] (0.8,5.7) node { $x_{1}$};	
		
	\end{tikzpicture}
	\label{fig2}
	\caption{}
\end{figure}

 We compute the integral
$$
A_{ij}^{\mu}:=\int_{\mathbb{R}^{\mathrm{B}_{d}}}e^{iX^{A}_{(x_{i},r_{i})}-iX^{A}_{(\theta \cdot x_{j},r_{j})}} d\mu_{A}
$$
via  a change of variables with respect to the map from $\mathbb{R}^{\mathrm{B}_{d}}$ to $\mathbb{R}^{2n}$ given by $(X^{A}_{(x_{1},r_{1})}, \ldots X^{A}_{(x_{n},r_{n})}, X^{A}_{(\theta \cdot x_{1},r_{1})},\ldots, X^{A}_{(\theta \cdot x_{n},r_{n})} )$,
and utilize the construction of $\mu_{A}$ from Theorem \ref{exigafie}, as well as Lemma \ref{findimreflpos} to conclude that the matrix $A_{ij}^{\mu}$
is positive definite. Since by construction the same holds if we replace $\mu_{A}$ with $\widetilde{\mu}_{A}$ and $X^{A}$ with $\widetilde{X}^{A}$, we conclude that $\widetilde{\mu}_{A}$ satisfies (\ref{mesreflosot})
for all $\Psi$ of form $z_{1}e^{i\widetilde{X}^{A}_{(x_{1},r_{1})}}+\cdots + z_{n}e^{i\widetilde{X}^{A}_{(x_{n},r_{n})}}$, where $z_{1}, \ldots ,z_{n}$ are arbitrary complex numbers.
		
Now let us consider the probability space $(C(\mathrm{B}_{d}^{+}), \mathcal{B}_{C(\mathrm{B}_{d}^{+})},\widetilde{\mu}_{A}^{+})$, where
$\widetilde{\mu}_{A}^{+}$ is the image of $\widetilde{\mu}_{A}$ under the restriction map 
$ R_{\mathrm{B}_{d}^{+}}: C(\mathrm{B}_{d}) \rightarrow C(\mathrm{B}_{d}^{+})$. 
Let $\widetilde{X}^{A,+}$ be the restriction of $\widetilde{X}^{A}$ to $C(\mathrm{B}_{d}^{+})$. By \cite[Corollary 7.12.2]{B} the $\mathbb{C}$-linear span of $\{e^{iF}\}_{F \in \mathrm{S}(\widetilde{X}^{A,+},\mathrm{B}_{d}^{+})} $ is dense in $L^{2}_{\mathbb{C}} (C(\mathrm{B}_{d}^{+}), \mathcal{B}_{C(\mathrm{B}_{d}^{+})},\widetilde{\mu}_{A}^{+}) $. It follows that any $\Psi \in \mathcal{M}(\mathrm{B}_{d}^{+})$ can be approximated in $L^2$ by a sequence of functions in the $\mathbb{C}$-linear span of $\{e^{iF}\}_{F \in \mathrm{S}(\widetilde{X}^{A},\mathrm{B}_{d})}$ and we conclude that $\widetilde{\mu}_{A}$ satisfies (\ref{mesreflosot}) for all $\Psi$ in $\mathcal{M}(\mathrm{B}_{d}^{+})$.
\end{proof}

Note that by the continuity of $\widetilde{X}^{A}$ and the measurable maximum theorem (cf. \cite[Theorem 18.19]{AB}) the suprema of $|\widetilde{X}^{A}|$ over compact subsets and hence over $\sigma$-compact subsets of $\mathrm{B}_{d}$ as well are Borel measurable. 
We fix $r>1$ and set $\mathrm{B}_{K,r}:= K\times (0,r]\subset
\mathrm{B}_{d}$ for every compact $K \subset \mathbb{R}^{d}$.
\begin{lemma}\label{supboule} $$\int_{C(\mathrm{B}_{d})} \sup_{b \in \mathrm{B}_{K,r}}| \widetilde{X}_{b}^{A}|d\widetilde{\mu}_{A} < \infty$$
\end{lemma}
\begin{proof}
We set	$\mathrm{B}^{n}_{K,r}:=  K\times [1/n,r]\subset
\mathrm{B}_{x,r} $ for $n=1,2,\ldots$ and denote by $\rho_{n}$ the canonical pseudo-metric associated to the restriction of $\widetilde{X}^{A}$ to $\mathrm{B}^{n}_{K,r}$ and by $N_{\rho_{n}}(\varepsilon)$  the minimum number of balls of radius $\varepsilon$ (with respect to $\rho_{n}$) whose union covers $\mathrm{B}^{n}_{X,r}$ (cf. Section \ref{prelimsect}). Now \cite[Theorem 1.3.3]{AT} implies the estimate
\begin{equation}\label{exptestion}
	\int \sup_{b \in \mathrm{B}^{n}_{K,r}} |\widetilde{X}^{A}|d\widetilde{\mu}_{A}\leq
	M \int_{0}^{\text{diam}(\mathrm{B}^{n}_{K,r})} N^{1/2}_{\rho_{n}}(\varepsilon)d \varepsilon
\end{equation}
for all $n$, where $M>0$ is a universal constant. Observe that the covariance function  defined in (\ref{defofmolcova}) is bounded on $\mathrm{B}_{K,r} \times \mathrm{B}_{K,r}$. It follows that 
the pseudo-metrics $\rho_{n}$ are uniformly bounded in the following sense: for every  pair $b_{1}, b_{2} \in \mathrm{B}_{K,r}$ the sequence  $\rho_{n}(b_{1}, b_{2})$ is defined for sufficiently large $n$ and is bounded. Thus we conclude that the functions $N_{\rho_{n}}(\varepsilon)$ are uniformly bounded in $n$ for every fixed $\varepsilon> 0$ which implies, as in the proof of Proposition \ref{ahascontpath}, that the left-hand side of (\ref{exptestion}) is bounded uniformly in $n$. Finally, since
$$\sup_{b \in \mathrm{B}_{K,r}}| \widetilde{X}_{b}^{A}| =  \liminf _{n \rightarrow \infty} \sup_{b \in \mathrm{B}^{n}_{K,r}} |\widetilde{X}_{b}^{A}|,$$
Fatou's lemma implies the desired result.
\end{proof}

Let us write $C_{0}(\mathrm{B}_{d})$ for the subset of $C(\mathrm{B}_{d})$ consisting of all functions whose restrictions to $\mathrm{B}_{x,r}$ are bounded for all compact $K \subset \mathbb{R}^{d}$ and all $r>0$. We equip  $C_{0}(\mathrm{B}_{d})$ with the locally convex topology generated by the family of semi-norms
$$\{\|\cdot \|_{K,r}:=\sup_{b \in \mathrm{B}_{K,r}} | \cdot | :  K \subset \mathbb{R}^{d}, r>1  \},$$
where $K$ is compact.
By Lemma \ref{supboule} one has
$$ \sup_{b \in \mathrm{B}_{K,r}}| \widetilde{X}_{b}^{A}(f)|< \infty $$
for almost all $f \in C(\mathrm{B}_{d})$, every $r>1$ and every compact $K \subset \mathbb{R}^{d}$, i.e. $\widetilde{X}^{A}$ has a modification with sample paths lying in $C_{0}(\mathrm{B}_{d})$. Since $C_{0}(\mathrm{B}_{d})$ is metrizable, Blackwell's theorem (see e.g. \cite[Corollary 8.6.8]{Co}) implies that the point evaluations generate the Borel $\sigma$-algebra on $C_{0}(\mathrm{B}_{d})$. Thus, proceeding exactly as in the end of Section \ref{prelimsect}, we can construct out of $\widetilde{X}^{A}$ and $\widetilde{\mu}_{A}$ a Borel probability measure $\bar{\mu}_{A}$ on $C_{0}(\mathrm{B}_{d})$. We write $\bar{X}^{A}$ for the canonical random field associated to it  and set $\mathcal{H}_{s}:=L^{2}_{\mathbb{C}} (C_{0}(\mathrm{B}_{d}), \bar{\mathcal{B}}_{\mathrm{B}_{d}},\bar{\mu}_{A})$.

\begin{lemma}\label{exisoflimpo} Assume that $A$ is bounded away from 0. Then for every $x \in \mathbb{R}^{d}$ and every sequence $r_{n}$ of positive numbers converging to 0, the sequence $\bar{X}^{A}_{b(x,r_{n})}$ has a subsequence converging pointwise on $C_{0}(\mathrm{B}_{d})$ as well as in the norm of $\mathcal{H}_{s}$ to a non-constant random variable. 
\end{lemma}
\begin{proof}
We first observe that one has
$$|\bar{X}^{A}_{b(x,r)}(f_{1}) -\bar{X}^{A}_{b(x,r)}(f_{2})|\leq 
\|f_{1} -f_{2}\|_{\{x\},r}$$
for all $x \in \mathbb{R}^{d}$, $r>1$, $f_{1},f_{2} \in C_{0}(\mathrm{B}_{d})$, 
which implies that the sequence $\bar{X}^{A}_{b(x,r_{n})}$ of functions on $C_{0}(\mathrm{B}_{d})$ is equicontinuous. Since it is also pointwise bounded, we conclude by a generalization of the  Arzel\`{a}-Ascoli theorem (cf. \cite[Chapter X, \S 2, No. 5, Corollary 1]{Bu}) that $\bar{X}^{A}_{b(x,r_{n})}$  contains a subsequence $\bar{X}^{A}_{b(x,r_{n_{i}})}$ converging pointwise to a continuous function on $C_{0}(\mathrm{B}_{d})$ which we denote by $\bar{X}^{A}_{b(x,0)}$.

Setting $r_{\text{max}}:=\sup_{n}r_{n}$ we note that 
$$|\bar{X}^{A}_{b(x,r_{n_{i}})}|\leq \sup_{b \in \mathrm{B}_{\{x\},r_{\text{max}}}} |\bar{X}^{A}_{b(x,r_{n_{i}})}|.$$
By Lemma \ref{supboule} and \cite[Theorem 2.1.3]{AT} the latter supremum is in $\mathcal{H}_{s}$, hence  $\bar{X}^{A}_{b(x,r_{n_{i}})}$ converges to $\bar{X}^{A}_{b(x,0)}$ in $\mathcal{H}_{s}$ by Lebesgue's dominated convergence theorem.

Finally, since $A$ is assumed to be bounded away from 0, the covariance function defined in (\ref{defofmolcova}) is bounded away from zero on $\mathrm{B}_{\{x\},r} \times \mathrm{B}_{\{x\},r}$. It follows that the norms $\|  \bar{X}^{A}_{b(x,r_{n_{i}})}  \|_{\mathcal{H}_{s}}$ are bounded away from 0, hence $\bar{X}^{A}_{b(x,0)}$ is non-zero, and by linearity, non-constant.
\end{proof}
We now take, as in Example \ref{exmolfref}, $A$ to be the operator $(I-\Delta)^{-1}$, fix a one-to-one continuous function $\phi: \mathbb{R} \rightarrow  \mathbb{R}$ and let $X^{\phi}$ be the transformation of  $\bar{X}^{(I-\Delta)^{-1}}$ via $\phi$, i.e., we set
$X^{\phi}=\bar{X}^{(I-\Delta)^{-1}}\circ \phi$. 
Setting $\mathcal{H}_{s}=L^{2}_{\mathbb{C}} (C_{0}(\mathrm{B}_{d}), \mathcal{B}_{C_{0}(\mathrm{B}_{d})},\bar{\mu}_{(I-\Delta)^{-1}})$ and
applying the construction from Section \ref{fromranfisec} to the random field $X^{\phi}$, we obtain a triple $(\mathcal{H}_{s},\mathfrak{A}_{X^{\phi}},1_{\mathcal{H}_{s}})$. In the following, note that $C_{0}(\mathrm{B}_{d})$ is equipped with a natural continuous action of $\mathrm{E}_{d}$.
\begin{theorem}
The triple  $(\mathcal{H}_{s},\mathfrak{A}_{X^{\phi}},1_{\mathcal{H}_{s}})$  is an algebraic Euclidean QFT over $\mathbb{R}^{d}$.
\end{theorem}
\begin{proof}
We shall verify that the triple $(\mathcal{H}_{s},\mathfrak{A}_{X^{\phi}},1_{\mathcal{H}_{s}})$  satisfies all assumptions of Theorem \ref{fromranfitoth}.

First, since the operator $(I-\Delta)^{-1}$ is $\mathrm{E}_{d}$-invariant, by Lemma \ref{lemtwoparo}(1) so is the measure $\widetilde{\mu}_{(I-\Delta)^{-1}}$. It follows that $\widetilde{\mu}_{(I-\Delta)^{-1}}$ is $\mathrm{E}_{d}$-invariant and  $X^{\phi}$ is $\mathrm{E}_{d}$-equivariant. By \cite[Corollary 7.12.2]{B} one has $\mathfrak{A}_{X^{\phi}}1_{\mathcal{H}_{s}}=\mathcal{H}_{s}$.

Second, we note that
$(I-\Delta)^{-1}$ is known to be reflection positive (see e.g. \cite[Theorem 7.10.1]{GJ}), hence by Lemma \ref{lemtwoparo}(2) the measure $\widetilde{\mu}_{(I-\Delta)^{-1}}$ is reflection positive in the sense of Definition \ref{mesdefrpos} which in turn implies that $X^{\phi}$ is reflection positive (cf. Definition \ref{rposforanf}) as well.

Finally, since $(I-\Delta)^{-1}$ is bounded away from 0, Lemma \ref{exisoflimpo} implies that for  every $x \in \mathbb{R}^{d}$ there exists a sequence of positive numbers $r_{n}$ decreasing to 0 such that $\bar{X}^{(I-\Delta)^{-1}}_{b(x,r_{n})}$ convergences pointwise  to a non-constant random variable. It follows that $e^{i\bar{X}^{\phi}_{b(x,r_{n})}}$ converges to a non-constant bounded random variable pointwise and, by dominated convergence, in $\mathcal{H}_{s}$ as well.
\end{proof}
\begin{remark}
Alternatively, one can consider $\bar{\mu}^{\phi}_{(I-\Delta)^{-1}}$, the image of $\bar{\mu}_{(I-\Delta)^{-1}}$ under the map given by
$$C_{0}(\mathrm{B}_{d}) \ni f  \mapsto \phi \circ f \in C_{0}(\mathrm{B}_{d}),$$ 
and the canonical random field on $(C_{0}(\mathrm{B}_{d}), \mathcal{B}_{C_{0}(\mathrm{B}_{d})},\bar{\mu}^{\phi}_{(I-\Delta)^{-1}})$   to obtain an algebraic Euclidean QFT equivalent to $(\mathcal{H}_{s},\mathfrak{A}_{X^{\phi}},1_{\mathcal{H}_{s}})$. Thus one can regard the triple $(\mathcal{H}_{s},\mathfrak{A}_{X^{\phi}},1_{\mathcal{H}_{s}})$ as being induced by the measure $\bar{\mu}^{\phi}_{(I-\Delta)^{-1}}$ which is in general non-Gaussian. 
\end{remark}


\begin{thebibliography}{9999}

\bibitem{AT} Adler, R. J., Taylor, J. E.: {\em Random Fields and Geometry}, Springer, New York, 2007.
\bibitem{AB} Aliprantis, C. D., Border, K. C.: {\em Infinite Dimesional Analysis}, third ed., Springer, New York, 2006.
\bibitem{A} Araki, H.: {\em Mathematical Theory of Quantum Fields}, Oxford University Press, Oxford, 1999.
\bibitem{B} Bogachev, V. I.: {\em Measure Theory}, Springer, New York, 2007.
\bibitem{Bu} Bourbaki, N.: {\em General topology. Chapters 5-10}, Springer, Berlin, 1989.
\bibitem{Co} Cohn, D. L.:  {\em Measure Theory}, Birkhauser, Boston, 1980.
\bibitem{Du} Dudley, R. M.: {\em Uniform Central Limit Theorems}, second ed., Cambridge University Press, Cambridge, 2014.
\bibitem{GJ} Glimm, J., Jaffe, A.: {\em Quantum Physics: A Functional Integral Point of View}, second ed., Springer, Berlin--New York, 1987.
\bibitem{HK} Haag, R., Kastler D.: An algebraic approach to quantum field theory, {\em J. Math. Phys.} {\bf 5} (1964), 848--861.
\bibitem{Ja} Janson, S: {\em Gaussian Hilbert Spaces}, Cambridge University Press, Cambridge, 1997.
\bibitem{Ka} Kahane, J.-P.: {\em Some Random Series of Functions}, second ed., Cambridge University Press, Cambridge, 1985.
\bibitem{KL} Klein, A., Landau, L.J.:  From the Euclidean group to the Poincar\'e group via Osterwalder-Schrader positivity, {\em Comm. Math. Phys.} {\bf 87} (1983), 469--484.
\bibitem{Kl} Klenke, A.: {\em Probability Theory}, second ed., Springer, London, 2014.
\bibitem{Sc} Schlingemann, D.: From Euclidean field theory to quantum field theory, {\em Rev. Math. Phys.} {\bf 11} (1999), 1151--1178.

\end{thebibliography}
\end{document}